\documentclass[11pt,twocolumn]{article}

\usepackage{times}

\usepackage[labelfont=bf]{caption}

\usepackage[top=0.5in,bottom=0.75in,left=0.5in,right=0.5in]{geometry}

\usepackage{url}

\usepackage{subfig}

\usepackage{graphicx}

\usepackage{amsmath}

\graphicspath{{figures/}{plots/}{.}}

\usepackage[numbers]{natbib}
\usepackage[bookmarksopen,bookmarksnumbered,colorlinks=true,allcolors=blue]{hyperref}

\usepackage{xcolor}

\usepackage[compact]{titlesec}

\usepackage{helvet}
\usepackage{sectsty}
\allsectionsfont{\sffamily}

\usepackage{mathptmx}   

\usepackage{flushend}

\usepackage{framed}

\usepackage[strict]{changepage}

\usepackage{booktabs}

\usepackage{adforn}
\usepackage{subfig}
\usepackage{type1cm}
\usepackage{lettrine}
\usepackage{enumitem}
\usepackage{url}
\usepackage{fontawesome5}

\newcommand{\systemname}{uxSense}
\newcommand{\Systemname}{uxSense}
\newcommand{\SYSTEMNAME}{uxSense}

\newcommand{\quotebullet}[1]{\noindent\adfrightarrowhead~\textit{#1}}

\newenvironment{widequote}%
  {\list{}{\leftmargin=0.1in\rightmargin=0.1in}\item[]}%
  {\endlist}

\newcommand{\ornamentbreak}{%
    \begin{center}
    \adfflatleafleft \quad\adfflatleafright
    \end{center}%
}

\begin{document}


\title{\sffamily uxSense: Supporting User Experience Analysis with Visualization and Computer Vision}

\author{\sffamily Andrea Batch,$^1$ Yipeng Ji,$^2$ Mingming Fan,$^3$ Jian Zhao,$^2$ and Niklas Elmqvist$^1$\\ 
\scriptsize\sffamily $^1$University of Maryland, College Park, MD, USA;\\
\scriptsize\sffamily $^2$University of Waterloo, ON, Canada;\\
\scriptsize\sffamily $^3$Hong Kong University of Science and Technology}

\date{\sffamily January 2023}

\maketitle

\begin{abstract}
    Analyzing user behavior from usability evaluation can be a challenging and time-consuming task, especially as the number of participants and the scale and complexity of the evaluation grows. 
    We propose \textsc{uxSense}, a visual analytics system using machine learning methods to extract user behavior from audio and video recordings as parallel time-stamped data streams.
    Our implementation draws on pattern recognition, computer vision, natural language processing, and machine learning to extract user sentiment, actions, posture, spoken words, and other features from such recordings. 
    These streams are visualized as parallel timelines in a web-based front-end, enabling the researcher to search, filter, and annotate data across time and space. 
    We present the results of a user study involving professional UX researchers evaluating user data using uxSense. 
    In fact, we used uxSense itself to evaluate their sessions.
\end{abstract}

\textbf{Keywords:} Visualization, visual analytics, evaluation, video analytics, machine learning, deep learning, computer vision.

\begin{figure*}[tbh]
    \centering
    \includegraphics[width=\linewidth]{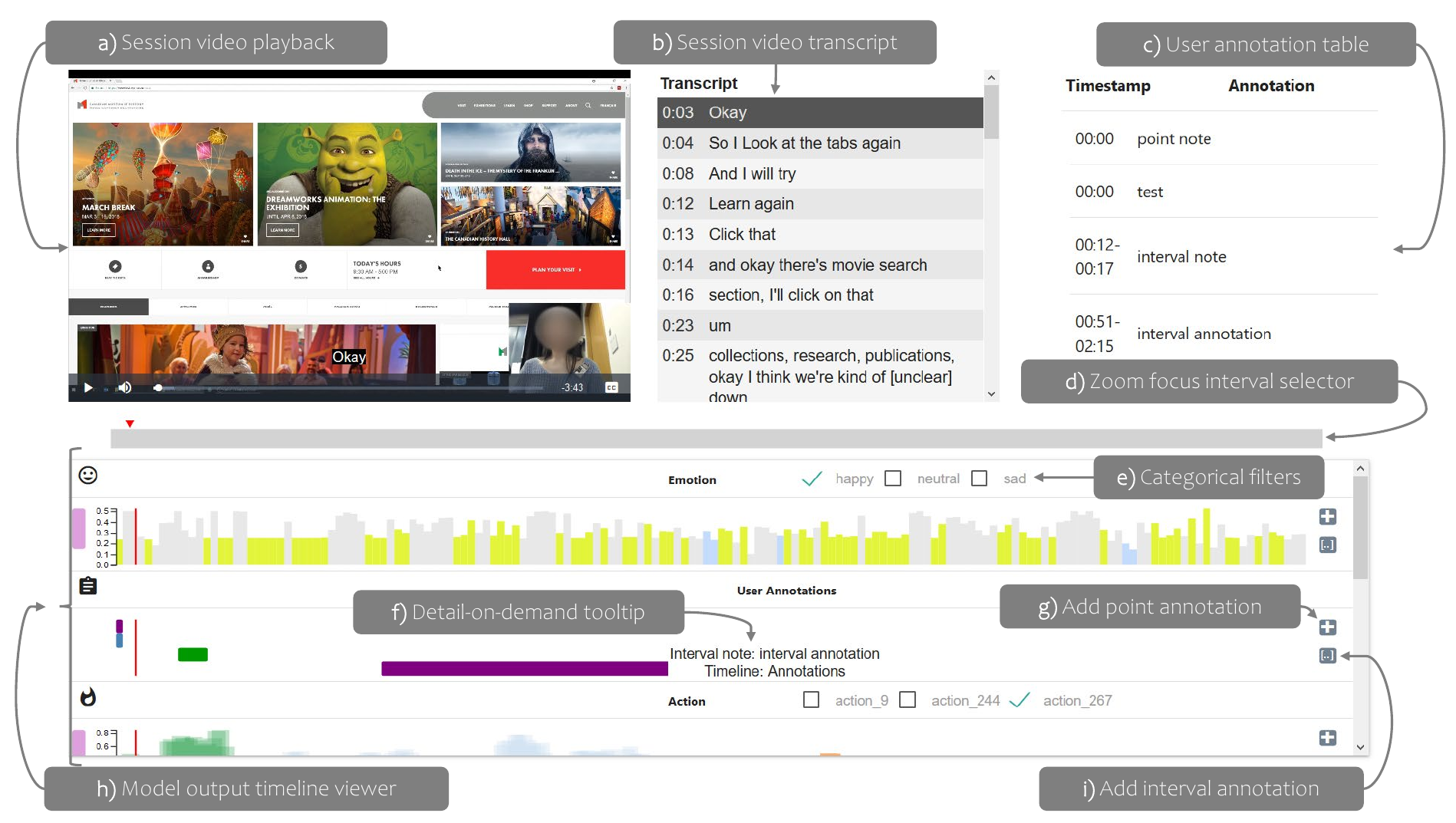}
    \caption{
        \textbf{Schematic overview.}
        Analysis interface in the \textsc{\systemname{}} web-based client (view reflects tutorial video). 
        \textbf{a)} \textit{Video playback}: View user session video, with or without captions.
        \textbf{b)} \textit{Session transcript}: View timestamped transcript of speech from video, and navigate video by clicking on line of text.
        \textbf{c)} \textit{User annotation Table}: View the text and timestamp of all annotations made by the user.
        \textbf{d)} \textit{Zoom focus}: Select, zoom, and pan whole extent of the video. Red arrow marker indicates current video time, while brushed region shows zoom extent in context of video duration.
        \textbf{e)} \textit{Categorical filters}: When selected, non-selected elements of the view are shown with low opacity.
        \textbf{f)} \textit{Details-on-demand}: Mouseover to get details of observation in model output at given time.
        \textbf{g)} \textit{Point annotation} and \textbf{i)}\textit{  Interval annotation}: Add an annotation corresponding to given timeline for either the video's current time (\textbf{g}) or the brushed interval range (\textbf{i}) with (\textbf{d}).
        \textbf{h)} \textit{Model output timeline viewer}: The timeline and user annotations are described in Section~\ref{sec:datamodel}.
    }
    \label{fig:system}
\end{figure*}

\section{Introduction}

We live in the age of the disappearing computer~\cite{Weiser1991} where most any gadget now involves computing technology and where the internet increasingly becomes integrated into everyday life.
As these devices and technologies reach an ever-expanding audience, the need for compelling user experiences increases.
This yields a growing need for usability, user research, and user experience (UX) professionals who can design, evaluate, and revise these interfaces.
Unfortunately, UX evaluation is a costly and time-consuming activity that does not scale with increased demand.

The bottleneck is most often the actual analysis of information, where a UX or usability professional often must review hours of video and audio recordings of different individuals who are interacting with a specific piece of technology, such as a smartphone, dishwasher, or website.
Further, a UX professional usually needs to attend to multiple behavioral signals (e.g., emotion, facial expression, body movement, and actions) simultaneously at a fast pace and combine them to make thorough and precise assessment of usability problems.
These signals are observed or inferred from the video and/or audio recordings, some of which may not always be salient to quick manual checking.
The above challenges make such video review a tedious and time-consuming task that often involves stepping forward the video a single frame at a time, repeatedly rewatching a critical segment, or manually coding the mood, actions, or body language of the participants.

We propose a method for extracting multi-modal features of human behavior from video and audio footage using machine learning (ML) to support UX and usability professionals in their analysis of user session data using interactive visualization.
Existing off-the-shelf tools for reviewing usability videos (e.g., UserTesting\footnote{\url{https://www.usertesting.com/}} and FullStory\footnote{\url{https://www.fullstory.com/}}) either just offer video playback functions or support extracting a few basic features (e.g., sentiment), lacking a holistic view of user interactions happening in the session.
They also do not provide a unified and interactive visual representation of multiple time-based metrics to support in-depth analysis.
While some systems (e.g., VA$^2$~\cite{Blascheck:2016} and CoUX~\cite{Jahangirzadeh2021}) concurrently visualize several sources of data in usability studies, they are still limited in their data features (i.e., eye-tracking, think-aloud, and interaction events) and are not flexible in integrating other time-based metrics.

Our prototype system---\textsc{\systemname{}}---validates the above method by providing a web-based interface to a computational backend that asynchronously runs a range of \textit{filters} to extract different types of data from one or several time-synced video streams uploaded by the user (Figure~\ref{fig:system}) as \textit{data streams}.
Filters in the \systemname{} system are designed as plugins that can be added or removed at run-time, each responsible for extracting one or more types of data, such as spoken language, gaze direction, arm gestures, head orientation, and even facial expressions.
Data streams recovered from each filter are shown in the visual interface as parallel time tracks, similar to the track-centric approach with a horizontal timeline used by video editing software.
Finally, the tool provides robust annotation features where the user can select events as well as intervals across the timelines and add notes for future analysis, collaborative iteration and review, and report generation.

To prove the feasibility of our method, we present results from an expert review involving five professional UX designers working at several large tech companies on the U.S.\ West Coast.
We asked these participants to use \systemname{} to analyze a think-aloud usability session performed on the Tableau desktop visualization tool.
To further showcase the utility of \systemname{}, we used the system itself to evaluate data streams recorded during these sessions.
Our findings show promise for our vision of ML to facilitate usability and UX testing, as well as for our \systemname{} prototype implementation.
While one of our participants wanted to know more about the accuracy of the models before she would trust it, the general sentiment among them was that the ideal use of such preprocessing filters could significantly ease their own daily work processes, or at least help them identify points of interest.
We close the paper by briefly speculating how future iterations of \systemname{} could also be used to evaluate mobile and ubiquitous analytics applications, where the participants walk around in physical space rather than being restricted to a workspace with a screen, mouse, and keyboard.

In summary, the primary contributions of this paper include: 
\begin{enumerate}
    \item A design framework for supporting UX research with interactive visual analytics of temporal UX metrics by incorporating off-the-shelf ML modeling of user video and audio;
    \item A system---\textsc{\systemname{}}---designed around our framework's design requirements; and
    \item Results from a user study evaluating the \systemname{} prototype.
\end{enumerate}

\section{Background}

In this section, we review the relevant work including visualization and human-computer interaction (HCI) evaluation, ML for extracting user behavior, and visual analytics to support UX studies.

\subsection{Data-driven Evaluation in HCI and Visualization}

In the domain of HCI and data visualization, some approaches to understanding factors influencing users' experiences and needs for systems involve constructing personas---representational archetypes of ``typical'' users and their daily lives~\cite{Holtzblatt2014}. 
This generally involves qualitative and ethnographic methods in which the researcher tracks, records, and interprets the users' daily activities in collaboration with the participant, reaching a shared understanding of the user's thought processes through interview and activity~\cite{Holtzblatt2014, Shilton2014}.
Alternative approaches exist in which events within the interface, such as mouse activity~\cite{Ahmed2007}, are used to develop ``data-driven personas''~\cite{ZhangA2016} for specific types of users.
Platforms for crowdsourcing experiments such as Amazon Mechanical Turk~\cite{Kittur2008} make the creation of these types of personas more manageable at larger scales.
In these evaluation methods, video and audio recordings tend to be a ubiquitous form of data to capture and analyze.

Video and audio recordings are nuanced data formats that are cheap to capture but expensive to analyze.
As a result, HCI and visualization research communities have already begun to shift toward the use of video and audio inputs as a revealed behavior dataset that is time-cost cheap and therefore scalable for the analysis of large user populations.
Systems for visualizing and analyzing visual and semantic features of cinematic films in the context of film studies exist. 
One example is VIAN~\cite{Halter2019}, which represents information about average frame color to the user, who can then manually segment the video with semantic annotations.
Kurzhals et al.~\cite{Kurzhals2016} introduce a system that uses the text of movie scripts to assign semantic labels to frames, which is graphically represented to the user along with motion and other visual frame information in an interactive dashboard that affords user annotation.
Pavel et al.~\cite{Pavel2014} present a system for automatically segmenting and summarizing lecture recordings and append them with crowdsourced transcripts.
QuickCut~\cite{Truong2016} allows for fast video editing and annotation to quickly transcribe, semantically match, and cut together audio annotations corresponding to timestamped clip segments.
Leake et al.~\cite{Leake2020} create a system for automatically generating audio-video slideshows using text and imagery from written articles.

Manual video annotation, such as using the ELAN tool~\cite{DBLP:conf/lrec/SloetjesW08}, has long been in existence, but recent years have also seen automatic methods that could be used for scalable evaluation based on video.
EgoScanning~\cite{DBLP:conf/chi/HiguchiYS17} processes first-person (egocentric) videos to detect important passages and adapts playback speed accordingly.
VidCrit~\cite{DBLP:conf/uist/PavelGHA16} compiles spoken, textual, and gestural feedback during video production into a visual interface for navigating annotations.
Finally, commercial services such as UserTesting and \texttt{Frame.io}\footnote{\url{https://www.frame.io/}} are based on video annotation and sensing, and while the former is focused on product evaluation and acceptance, it still does not provide sophisticated visual analytics tools to study this data.
Compared to these efforts, \systemname{} provides an extensible list of AI/ML/CV filters that extract various features (e.g., user actions, emotions, speech rate, etc.) from the audio and video recordings, thus providing a holistic view of user behavior.

Finally, in EduSense~\cite{DBLP:journals/imwut/AhujaKXVXZTHOA19}, automatic video annotation is applied at scale to classroom sensing, allowing an array of commodity cameras to capture previously latent metrics about learning environments.
We tend to think of the \systemname{} system as having a similar scope: it captures many previously intangible metrics about user experience from video recordings into a platform that enables UX researcher to explore and analyze this data.

\subsection{Characterizing Users with Machine Learning}

The HCI community has been harnessing artificial neural networks (ANNs), including recurrent and convolutional neural networks (RNNs and CNNs), for evaluating user behavior.
Examples include discovering speech patterns~\cite{fan2019vista}, identifying gesture~\cite{Marques2015, Ramakant2015, Tompson2014} and gaze~\cite{Mukherjee2015, Zhang2015}, classifying user emotion and facial expression~\cite{Harezlak2014, Meng2016, Suja2015}, and detecting characteristics of the user, such as gender~\cite{Wolfshaar2015},\footnote{We note that this approach, like many similar projects conceived with little thought to their sociotechnical impact, are a highly questionable practice.} by constructing and implementing neural network architecture.
The visualization community has also made contributions to the toolkit of methods used in evaluating user video, logs, transcripts, and other qualitative data~\cite{Chandrasegaran2017}, as well as user gesture analysis~\cite{Kerdvibulvech2007}. 

In the ML community there has been more than a decade's worth of research exploring methods for action classification~\cite{Laptev2008, Niebles2007}, motion and path prediction\cite{Ma2017}, eye tracking~\cite{Krafka2016}, and gesture detection~\cite{Molchanov2016}.
While there have been a few position papers~\cite{Cambria2013} and more serious studies~\cite{Heng2014} advocating for a closer relationship between the HCI and machine intelligence communities, the current body of literature on the subject is still somewhat sparse. 

Our work fills in these gaps by applying suitable ML techniques to the analysis of video and audio recordings in the domain of UX evaluation in HCI and presenting the results interactively for in-depth sensemaking.
Specifically, \systemname{} leverages the available body of research in speech analysis, computer vision, and machine intelligence for characterizing audio and video inputs into different types of information, i.e., data streams. 
Each of the data streams is generated by independent functional units that we call filters, and then displayed with multiple coordinated timeline visualizations.

\subsection{Visual Analytics to Facilitate UX Evaluation}
\label{sec:va-ux}

As ML continues demonstrating its potential, qualitative researchers are becoming increasingly interested in adopting ML into their analysis flows.
However, challenges quickly surface.
First, although traditional classification and clustering ML methods are helpful for generating additional labels to inform analysis, these labels alone are often not sufficient for addressing HCI research problems.
Instead, HCI researchers need to leverage their skills to make sense of the ML-generated labels to gain a deeper and more nuanced understanding of the data. 
Second, many ML methods require a significant amount of data to optimize parameters and thus have limited accuracy when dealing with small-scale yet semantically-rich human-behavior data.
This has inspired new methods, such as visualization, for better integrating ML into qualitative workflows.

One line of research is to support qualitative coding, which is a powerful yet labor-intensive method.  
Felix et al.\ designed a visual data analysis tool that integrates unsupervised learning methods to provide suggestions to help researchers progressively code a large corpus of texts~\cite{Felix2018Exploratory}.
Another challenge that qualitative researchers often face is to resolve conflicts among researchers when analyzing qualitative data.  
Drouhard et al.\ designed a tool, Aeonium, that identifies potential conflicts in codes created by different coders using ML and highlights the conflicts to facilitate coders to spot their disagreements and resolve conflicts efficiently~\cite{Drouhard:2017}. 

Another line of research is to support the analysis of user interaction data to uncover users' intentions and reasoning processes.
Both low-level user inputs (e.g., mouse clicks, drags, key presses~\cite{robinson2006re, Gotz2008}) and high-level graphical structures of user interactions~\cite{Heer2008} are captured and visualized to help researchers make sense of their analytic activity. 
Moreover, eye-tracking data have also been visualized to help researchers analyze user interactions and even predict user intent~\cite{Blascheck2017, silva2018leveraging}. 
Furthermore, researchers have investigated user-generated annotations and developed visual interfaces to uncover hidden sensemaking patterns~\cite{Zhao2017a, Zhao2017b}.
In addition to using proxy data (e.g., mouse events, eye-tracking data) and manual provenance (e.g., user-generated annotations), researchers have recently begun to investigate think-aloud data, which are generated by asking users to verbalize their thought processes while working on a task, to better understand their hidden thinking process. 
Think-aloud data have been used to understand analysts' reasoning processes~\cite{Lipford2010, Dou2009} as well as users' interactions~\cite{fan2020practices}.
VA$^2$ visualizes think-aloud, interaction, and eye movement data to facilitate the analysis of multiple concurrent evaluation results~\cite{Blascheck:2016}.
However, VA$^2$ only supports these three data streams, in raw forms, with a specific visualization design, which lacks a more holistic view of the user behaviors that can be characterized by other data or derived features.
Recently, ML models have been employed to predict usability problems of think-aloud sessions (e.g., based on users' speech, verbalization, and scrolling patterns), and the ML predictions are further visualized in timelines to support individual~\cite{fan2019vista} and collaborative~\cite{Jahangirzadeh2021} UX evaluation.
In addition, analytical technologies can detect user moods and facial expressions to facilitate UX evaluation~\cite{staiano2012ux_mate, tan2014inferring, munim2017towards, da2019uxmood}.

Inspired by prior work, we extend this line of research by considering a wider range of modalities of data extracted from video and audio footage that are indicative of users' experiences (speech rate, transcripts, gaze direction, facial expressions, semantic actions).
These time-stamped data streams can then be combined into a flexible and extensible timeline visualization panel to enable detailed analysis of user behaviors using visual analytics.

\begin{table*}[tbh]
    \centering
    \caption{List of the current filters implemented in \systemname{}.}
    \begin{tabular}{llllp{6.5cm}}
    
        \toprule
        \textbf{Name} & \textbf{Filter Description} & \textbf{Data Stream} & \textbf{Model} & \textbf{Justification} \\
        \midrule

        VisTA & Speech transcription & Text, speech rate & \cite{fan2019vista,fan2022auto} & Transcripts are standard for usability analysis; user speech tends to slow down when users encounter a usability issue.\\ 
        
        CoUX & Audio & Pitch & \cite{Jahangirzadeh2021} & User speech tends to change pitch when users encounter a usability issue.\\
                
        VideoPose3D & Track the user's position & Continuous vectors & \cite{Pavllo2019} & User position reflects embodiment and intention for MR.\\
        
        E-Divisive & Joint angle-based intervals & Frame intervals & \cite{Batch2018gesture, Matteson2014} & User pose reflects embodiment and intention for MR.\\
        
        Kinetics-I3D & Semantic action classification & Action probabilities & \cite{Carreira2017}, \cite{Batch2019mlui} & User actions reflect embodiment and intention for MR.\\
        
        face-classification & Real-time emotion classification & Emotion probabilities & \cite{Arriaga2017} & User mood often reflects their experience using a tool.\\
        
        \bottomrule
    \end{tabular}
    \label{tab:filters}
\end{table*}

\section{Extracting Behavior from Video}
\label{sec:framework}

Our framework is based on visualizing time-stamped feature streams extracted from audio and video using entirely automatic methods from computer vision, machine learning, and signal processing.
This semantic information is presented to the UX researcher in a visual timeline interface that supports user session video evaluation, manual annotation, and report generation.
The purpose for our framework is to enable an analyst to triangulate multiple data sources with their own knowledge and experience to understand UX issues in a more accurate and efficient manner.
We discuss our design constraints, data model, and applications below.

\subsection{Method: Requirements Analysis}

User experience is a data-driven discipline based on quantifiable and measurable data~\cite{Albert2013}.
While roles in UX are ill-defined, we can distinguish between two main types: \textit{UX researchers}, who take a proactive role in understanding \textit{what} users want, and \textit{UX designers}, who are more concerned about \textit{how} to implement the product.
Our goal is to design a visual analytics tool that serves both roles.

In curating our design requirements and data features below, we draw on both industrial and academic resources documenting UX practices as well as the needs of UX designers and researchers:

\begin{itemize}
    \item \textbf{Industry UX:} Albert and Tullis~\cite{Albert2013} detail the quantitative metrics that UX research and design needs for effective analysis.
    Nunnally and Farkas~\cite{Nunnally2016} also discuss some of the more qualitative metrics involved.
    Sauro and Lewis~\cite{Sauro2016} review the formal statistical and mathematical methods commonly used

    \item \textbf{Academic UX:} Section~\ref{sec:va-ux} gives an overview of current academic tools for UX R\&D.
    In deriving our requirements, we draw upon our past work on supporting think-aloud sessions~\cite{fan2019vista,fan2022auto}, usability testing~\cite{Fan2019TOCHI}, and UX research~\cite{Fan2021CHI}.
\end{itemize}

While we would argue that the requirements listed below are the central one to quantitative and qualitative UX analysis, there are many additional requirements and criteria in the UX discipline.
Exhaustively listing these is beyond the scope of this paper.

\subsection{Design Requirements}

Supporting UX research means keeping UX research workflows central.
We propose the following interrelated requirements, each reflecting an important step in the UX research workflows:

\begin{itemize}

    \item[\faIcon{key}]
    \textbf{R1 - Semantic key point detection:}
    UX research workflows frequently involve identifying semantically significant or pivotal moments in user sessions upon reviewing session data.
    As such, the UX system should reduce the time cost of identifying important moments in user sessions.

    \item[\faIcon{layer-group}] \textbf{R2 - User-defined segment classification:}
    Identifying key points (R1) is often followed by classifying or tagging segments of the user session based on recurring or novel patterns in the data.
    The system should support constructing qualitative classification frameworks for session analysis.

    \item[\faIcon{pen-fancy}]\textbf{R3 - Annotation:} 
    Once segments have been identified (R1) and tagged (R2), they will be annotated by the analyst.

    \item[\faIcon{file-alt}] \textbf{R4 - Summary Report Generation:} 
    Finally, these identified (R1), tagged (R2), and annotated (R3) data streams should be summarized by the system as reports and figures.

\end{itemize}

\subsection{Data Features}

We model the following features from user data:

\begin{itemize}

    \item[\faIcon{water}]\textbf{F1 - Multiple concurrent streams:} 
    UX experiments often include multiple metrics measured over time, such as head position, physical location, activity level, speech, mood, etc.
    
    \item[\faIcon{angry}]\textbf{F2 - Facial expressions:} The ability to track facial expression may yield an insight into the user's emotional state.
    
    \item[\faIcon{comment}]\textbf{F3 - Speech:} User session video typically includes audio that captures dialogue between the researcher and the participant, as well as think-aloud or pair analytics procedures that involve the participant verbally externalizing their sensemaking process. 
    An evaluation system should take advantage of this information by generating a transcript from speech, visually representing features of the audio signal, and using it in computer vision models for predicting activity.

\end{itemize}

\subsection{Data Model and Filters}
\label{sec:datamodel}

Due to the computational time costs of predicting semantic features of video data, we anticipate a data model consisting of uploaded video files as input, with server-side computation processing occurring asynchronously over a brief period of time before model output is accessible to the client.
However, the design ideal would be in minimizing the latency between video input to model output.
For this reason, we have opted to use real-time implementations (e.g., using a real-time emotion classification framework~\cite{Arriaga2017}) where doing so would not heavily compromise the accuracy of our output.
Table~\ref{tab:filters} provides a cross-reference of the models (or ``filters'') involved in our present pipeline.

Model output from all filters is represented as data streams in synchronous timelines that can be flexibly manipulated.
These data streams can be linked to annotations in what may be called a ``human-in-the-loop'' stage of the pipeline.
Following this human-in-the-loop evaluation, the final output of the pipeline is generated as report figures.
The transcript also appears again at the final stage of the process as part of the micro-document output.

\subsection{Practical Considerations}

In addition to the above design requirements, it is also possible to combine our sensing mechanisms with data from clickstreams, IR motion tracking, and sociometric badges.
The parallel timelines in uxSense can accommodate any form of data.

The facial emotion classification network, which relies on finer features of the video subject's face, is calculated concurrently and asynchronously over fixed-width intervals using a rolling window.
Because of this reliance on finer features, we use 30 FPS video input for emotion classification.
However, this does not greatly affect model performance, since we have chosen a real-time model~\cite{Arriaga2017} for deriving the emotion labels.

Concurrently, the audio from the video input is used to derive transcripts, speech rate, and pitch~\cite{fan2019vista}.
It can be used for captions.

Finally, more specialized filters can be added as new plugins depending on context.
For example, it might be useful to track the 2D mouse pointer for desktop applications, whereas a Virtual Reality application could benefit from capturing the user's physical location, pose, and limb movements in 3D (see Section~\ref{sec:beyond-desktop}).

\section{The \SYSTEMNAME{} System}
\label{sec:system}

\Systemname{} is a client/server system that implements our framework for extracting user behavior from video (Section~\ref{sec:framework}) with a computational and storage backend and a web-based interactive frontend.
In broad terms, \systemname{} supports analysis of both qualitative and quantitative user experience through a variety of temporal visualization techniques for continuous (time series) and discrete (event) timelines representing \textit{feature extraction filters} (or just filters)---the products of our models (Table~\ref{tab:filters})---to highlight segments of interest.
Because many video analytics algorithms require significant processing to complete, \systemname{} is based on an asynchronous steering workflow where the user can shut down the interface while the video is processed on the server.
Results are streamed back to the client as it is made available.
The tool provides an interactive visual analysis interface for viewing multiple structured data streams, annotating them, and generating reports. 

\subsection{Overall Workflow}
\label{sec:workflow}

The main \systemname{} workflow is entirely asynchronous; any step of the following (all conducted using the web-based interface) process can be visited at any point (Figure~\ref{fig:teaser}):

\begin{enumerate}
    \item[(1)]\textbf{Upload} one (or more) video or audio recordings to the server.
    \item[(2)]\textbf{Start} asynchronous computation of the available list of filters.
    \item[(3)]\textbf{Monitor} computation progress in real time, or    
    \item[(4)]\textbf{Close down} the interface while computation continues.
    \item[(5)]\textbf{Analyze} the data as it becomes available.
    \item[(6)]\textbf{Generate} reports from data analysis.
\end{enumerate}

Each step is associated with a specific project.
The backend constantly runs computations while there are still recordings to process and filters that have not been executed.
Uploading new footage will schedule new computation for the unprocessed video.
Results are streamed dynamically to the analysis interface.

\subsection{Feature Extraction Filters}

The basic building block of \systemname{} is the \textit{feature extraction filter}, an algorithm that is capable of processing sequential video data $v_i$ and generating a corresponding sequence of extracted features $d_i$ (or \textit{data streams}), e.g.\ $f_{filter} : (v_1, \ldots, v_t) \rightarrow (d_1, \ldots, d_t)$.
Video frames consist of both imagery and audio, and extracted features are derived from any of these (or both).
For example, a typical feature extraction filter may track the position of a person in the frame over time, the direction of their gaze, and their perceived fatigue.
Some features are continuous, such as the user's head direction, whereas others are discrete, such as time intervals when a person is pointing or forming another gesture.
In addition to the sequential frame data, filters also maintain summary and aggregate data relevant to the tracked feature, such as the user's cumulative movement, their activity level, individual gestures, etc.

Our prototype \systemname{} implementation provides an initial library of feature extraction filters (see Table~\ref{tab:filters}). 
Many practical computer vision models track multiple features at once, such as the user's pose and a semantic label overlaid on top of the video playback.
However, the \systemname{} design philosophy is to provide feature extraction filters such that the user can rearrange, hide, and reveal streams based on what they deem most important, relevant, or revealing for their analysis.
This facilitates a more semantically meaningful configuration of which filters to include based on the research question and data being evaluated.

\subsection{Analysis Interface}

The \systemname{} analysis interface provides a mechanism for viewing and comparing feature data streams for one or multiple video recordings in parallel and time-synchronized \textit{tracks} akin to video editing software.
Figure~\ref{fig:system} shows a schematic overview of this interface. 
The three main elements of the analysis interface are the footage view, the text view, and the timelines (or tracks).
A common \textit{time indicator} on the \textit{track} governs which frame of the current footage is being viewed.
The \textit{footage view} shows that frame from the currently selected recording.
The \textit{text view} displays timestamped textual information about the video from the transcript and from the user's own annotations, and can be clicked to navigate to different points of interest throughout the video.

\begin{figure}[htb]
    \centering
    \includegraphics[width=\linewidth]{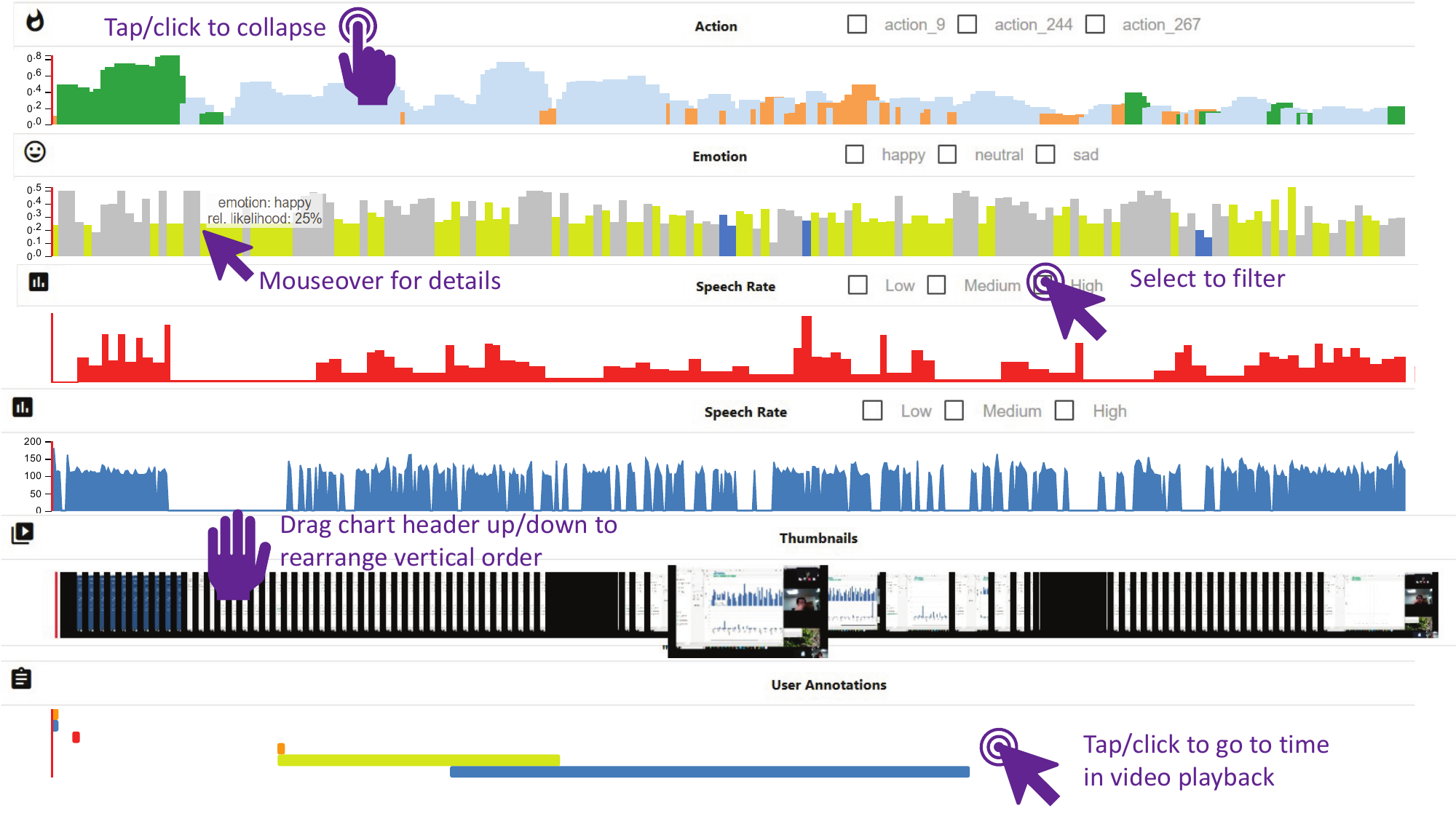}
    \caption{\textbf{Timeline View.}
    This view acts as both a video scrubber to select current time and an interactive representation of the filter output.
    Mousing over an element shows text details of the timeline at the current frame. 
    Filtering using the checkboxes on the header highlights all observations meeting the filter criteria by making all other observations semi-transparent.
    Clicking on the timeline navigates the video to the selected timestamp.}
    \label{fig:timelines}
\end{figure}

Each of the \textit{multiple concurrent data streams} (F1) are visualized depending on data type (from top to bottom in Figure~\ref{fig:timelines}):

\begin{itemize}[itemsep=0.0mm]
    
    \item\textbf{Action Predictions:} A plot of discrete events (using hues) based on action labels assigned to video segments over time, with prediction probability represented with rectangle height.
    Detecting actions support \textit{semantic key point detection} (R1).
    
    \item\textbf{Emotion Prediction}: \textit{Facial expression} (F2) emotion classification labels and prediction probabilities~\cite{Arriaga2017} are presented, also for supporting \textit{semantic key point detection} (R1).
    
    \item \textbf{Speech Rate}: \textit{Speech} (F3) is calculated using the word frequency over fixed time intervals using speech-to-text model output~\cite{fan2019vista}; we argue that this, too, supports a level of \textit{semantic key point identification} (R1) for the user session.
    
    \item \textbf{Pitch}: Audio signal pitch (see Section~\ref{sec:impnotes}).
    
    \item \textbf{Thumbnails}: Thumbnails with the moused-over video timestamp frame shown in relief larger than the others, which are dynamically repositioned using Cartesian fisheye distortion.
    
    \item \textbf{Annotations}: In support of a combination of \textit{semantic key point detection} (R1), \textit{user-defined segment classification} (R2), and \textit{annotation} (R3), the user's own annotations are visualized as a step function, with step colors signifying the annotation's timeline, and mouseover details showing the annotation and name of the corresponding timeline.

\end{itemize}

\begin{figure}[tbh]
    \centering
    \includegraphics[width=\linewidth]{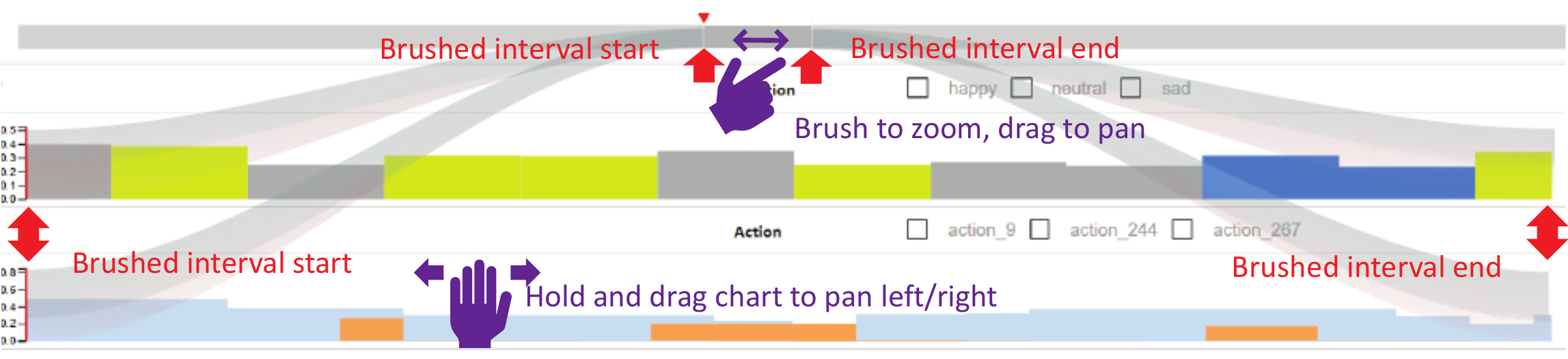}
    \caption{\textbf{Focus-brushing.}
    This feature selects an interval of the video, zooms all timelines, and allows the user to drag either the timelines or the selected rectangle to pan through the data.}
    \label{fig:zoompan}
\end{figure}

Beyond the time-marker based track view of each feature data stream, the user can zoom in on a point of interest by brushing the focus interval selection (Figure~\ref{fig:zoompan}).
Once zoomed, the timelines can be dragged to pan through the video and all of the timelines.
To support \textit{annotation} (R3)---and \textit{user-defined segment classification} (R2) by way of \textit{annotation} (R3)---the brushed interval of the video can be annotated as a range, or the user can opt to annotate a single point of the video as they code user behavior. 

\subsection{Micro-Report Generation with \systemname}
\label{sec:annotlettesystem}

\begin{figure}[htb]
    \centering
    \includegraphics[width=\linewidth]{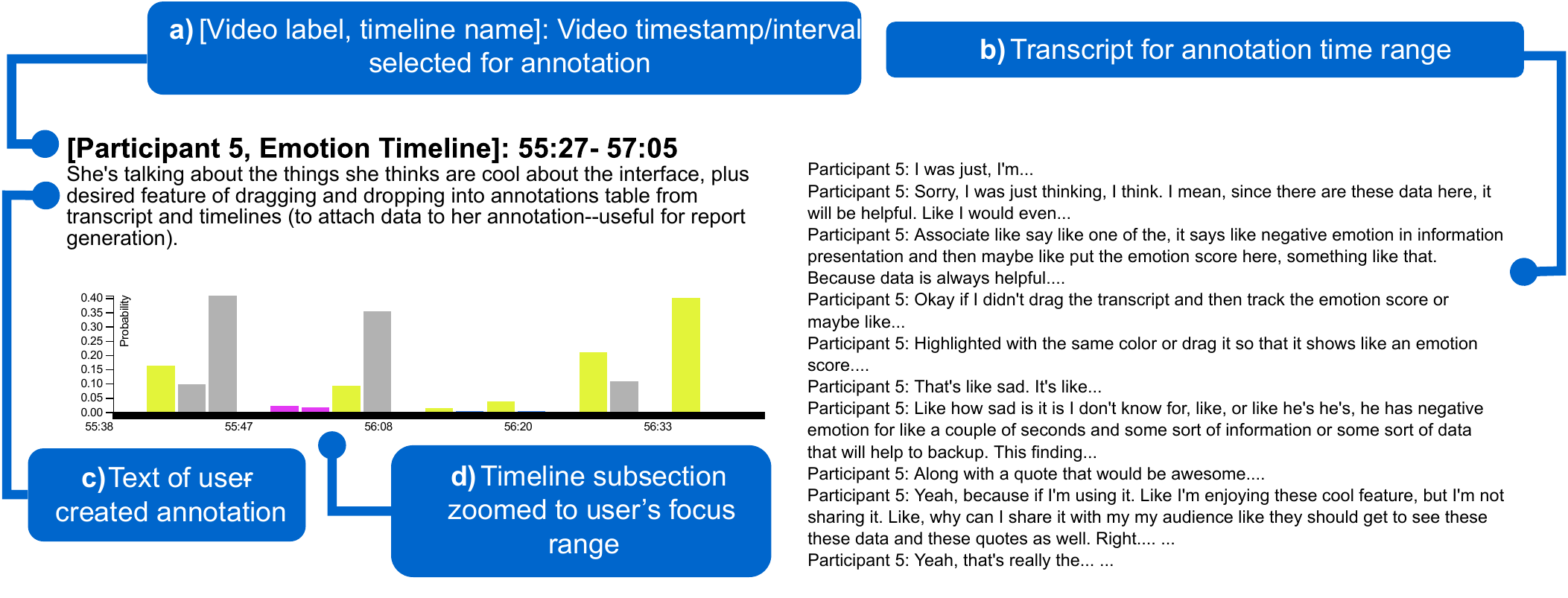}
    \caption{\textbf{Annotlette example.} An annotlette generated by \systemname{} during our own evaluation of user sessions in which UX professionals used \systemname{}.
    \textbf{a)} Metadata regarding the annotation.
    \textbf{b)} Transcript from the period selected for annotation.
    \textbf{c)} The annotation text created by the user.
    \textbf{d)} A zoomed view of the timeline associated with the annotation for the selected period.}
    \label{fig:annotlette_schematic}
\end{figure}

The final destination of \systemname{}---the concluding step in our workflow (Section~\ref{sec:workflow})---is to generate figures that can be used to report on user behavior in an academic paper, internal memo, or industry whitepaper.
The report generation functionality of \systemname{} creates a small vector graphic document for each individual annotation created by the user that we have named ``annotlettes'' (Figure~\ref{fig:annotlette_schematic}) that link the relevant timeline, transcript, and user annotation for the notes the user has created in the analysis.
The annotlette feature was added after we completed our user study based on participant feedback and our observations (see Section~\ref{sec:redesign}) evaluated through the lens of our design requirements; it directly links \textit{annotation} (R3) with \textit{summary report generation} (R4).

\begin{itemize}
    \item\textbf{Timeline Chunk:} A zoomed view of the annotated timeline (Figure~\ref{fig:annotlette_schematic}d) is used to represent the data stream segment; and
    \item\textbf{Transcript Snippet:} A static version of the transcript (Figure~\ref{fig:annotlette_schematic}b) during the selected time period; and
    \item\textbf{Annotation:} The user's annotation text, with the metadata associated with it formatted as a header (Figure~\ref{fig:annotlette_schematic}a\&c). 
\end{itemize}

\subsection{Implementation Notes}
\label{sec:impnotes}

We implemented \systemname{} as a Node.js\footnote{\url{https://nodejs.org/}} application with several of the server-side components implemented in Python and R that are spawned from the Node.js process.
Individual filters used a combination of freely available model implementations; see Table~\ref{tab:filters} for details.
For audio analysis, we used Praat\footnote{\url{http://www.praat.org/}} to extract audio features.
We also use the Google Cloud Speech-to-Text API\footnote{\url{https://cloud.google.com/speech-to-text}} to transcribe audio to text. 
The web client was implemented in JavaScript, HTML5, and CSS. 
We used HTML5 video and canvas for video playback, and D3.js~\cite{Bostock2011} for the visualization components.
Transcription is implemented using videojs-transcript.\footnote{\url{https://github.com/walsh9/videojs-transcript}}

Since some models provide multiple features in the same computation, we used a server-side caching scheme where different filters that rely on the same model look up previously computed data instead of rerunning the same analysis from scratch.
For example, a head-tracking filter that uses a full-body 3D tracking model to merely extract the user's gaze direction would store the recovered full 3D skeleton of the user in a local file.
If another filter was introduced that relied on the same model (e.g., determining the user's position in 3D space), that filter could merely look up the previously stored data instead of rerunning the same model.

Our framework is Open Source and can be accessed on GitHub at \url{https://github.com/DreaJulca/uxsense}.
Furthermore, as we are keen to make this technology available to other researchers working in this area, we also distribute prebuilt software packages on the \systemname{} GitHub website to facilitate dissemination.


\begin{figure*}[tbh]
  \centering
  \resizebox{\textwidth}{!}{\includegraphics[width=\textwidth]{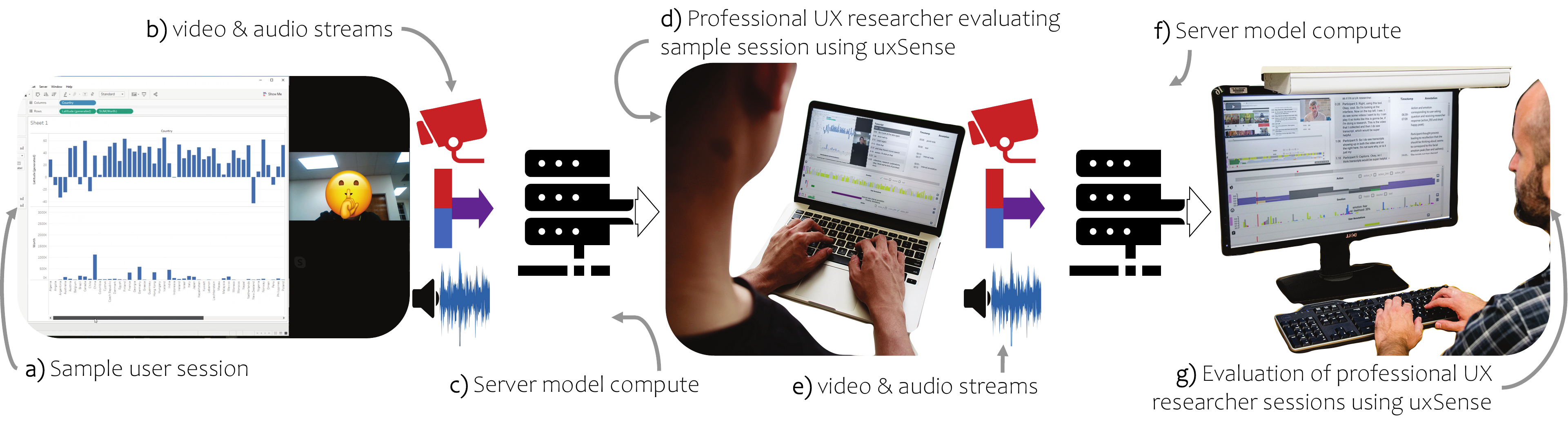}}
  \caption{\textbf{Evaluation pipeline.}
  Workflow in the \textsc{\systemname} prototype system for extracting user behavior from video footage using deep learning to support in-depth and advanced analysis of participant performance in user studies.
  We use \systemname{} to evaluate user sessions in which professional UX researchers use \systemname{} to evaluate a sample Tableau user session. 
  \textbf{a)} sample user session with commercial visual analytics tool (Tableau Public); 
  \textbf{b)} sample user session video and audio streams; 
  \textbf{c)} server compute of sample user session data: Video pose-estimate-based temporal segmentation, video emotion and action classification, speech detection and audio signal processing; 
  \textbf{d)} evaluation user sessions with professional UX researchers using \systemname{} interface with sample session video, model output; 
  \textbf{e)} UX researcher user session video and audio streams;
  \textbf{f)} server compute of video and audio models using professional UX researcher session data;
  \textbf{g)} authors’ evaluation of professional UX researcher user sessions using \systemname{}.
    }
  \label{fig:teaser}
\end{figure*}

\section{Expert UX Designer Review}
\label{sec:evaluation}

Our evaluation involved several professional UX researchers from large tech companies in the United States.
The study process was not only a means for evaluating our participants' responses, but was also an opportunity for us to use \systemname{} itself to analyze our users' evaluation of user session footage using \systemname{}.
Figure~\ref{fig:teaser} shows the study workflow.
Due to the limited availability of our participants, we chose to have participants use \systemname{} for an in-depth case study rather than conduct a comparative analysis between \systemname{} and alternative tools or baselines.

Prior to running actual tasks, we piloted the study with a single HCI master's student who was not familiar with the system. 
This pilot session enabled us to identify and troubleshoot issues with the software, as well as refine the testing protocol.

\subsection{Participants}

All participants were women who held the job title ``UX Designer;'' one was a ``senior UX designer.''
Two participants had 5 years of professional experience, two had 2-3 years of professional experience, and one had 0.5 years of professional experience (in addition to relevant graduate school experience).
Two participants held Ph.D.s in Information Science, one held a Ph.D.\ in HCI, one participant held a Master's degree in Product Innovation, and one participant held a Bachelor's degree in Media and Advertising.

When asked about their typical evaluation process, participants responded with the following descriptions:

    \quotebullet{``I normally video/audio record the session, take notes during the session and sometime collect data about the tasks we ask people to do during the session.
    User sessions are typically semi-structured with some tasks and open feedback.
    I normally code the data based on themes in spreadsheets and also create video clips that demonstrate the key themes.''}
    
    \quotebullet{``There are different UX research methods and they are conducted differently.
    The most common ones are usability test, interview, survey.
    In usability test, I mark the success of each task, quantify some of the useful actions, such as user errors, user habits, user preferences and quotes.
    I write down their pain points and extract themes from there.''}
    
    \quotebullet{``[I]nterview, survey, observation, usability testing.''}

    \quotebullet{``[B]ehavioral and perception data [...] or self-reported response about their workflow, interview, survey, diary study.''}\\

Participants reported frequently using the following categories of tools for evaluating user data:

\begin{itemize}
    \item \textbf{Video conferencing}: Hangouts, Zoom, and GoToMeeting.
    
    \item \textbf{Spreadsheets}: Microsoft Excel and Google Sheets.
    
    \item \textbf{Programming languages and extensions} for working with quantitative data and for NLP: R, Python, and internal tools.
    
    \item \textbf{Survey tools}: Validately, Qualtrics, and QuickTimer.
\end{itemize}

\subsection{Apparatus}

\Systemname{} was hosted on an institutional server running Linux (Ubuntu version 16.04.1) with an Intel Xeon CPU E5-1650 v3 (3.50GHz).
Because our system's backend pipeline involves a processing time longer than real-time and depends on GPU support for efficient model output, the model output data accessed by our users was pre-generated prior to the beginning of their session on a laptop running Windows version 10.0.18362 with an Intel Core i7-8750H CPU (2.2GHz) and a Nvidia GeForce GTX 1060 GPU (8GB RAM).
Users were able to access the system remotely. 

User interactions with the system and video playback activities were asynchronously posted to the server. 
Session video and audio activity was recorded using Zoom's Enterprise cloud service, which also generated transcripts of the user sessions that were evaluated both outside and within \systemname{}.
Video of the users' faces and audio of their voices during the session was recorded using their own webcams and microphones.
The model output for the video and data collected during the session was generated using the same laptop that was used to pre-generate data for the user sessions.

\subsection{Tasks and Procedures}
\label{sec:procedure}

Participant activity involved a user session 60-75 minutes long followed by a Google Forms survey that they were asked to fill out at their earliest convenience.
At the end of the session, participants were paid US\$50/hour for their time, with the post-session survey being compensated for in advance as a half-hour's worth of work. 

Participants were asked to perform an open-ended exploration of the \systemname{} interface features in a pre-activity training phase.
Once they were done with their exploration, the researcher informed the participant of any features not discovered during exploration, and answered their questions about the system.
During the second stage of the user session, the participants were tasked with identifying problems with the user's experience in an 11.67 minute-long video of a novice Tableau user exploring a dataset they hadn't seen before.
They were told that they would need to give a brief summary of their observations at the end of the session.
After this activity ended, participants were paid via Venmo and emailed a link to the post-session survey. 
All participants who completed the session also completed the post-session survey.

Participants were asked to think aloud during all stages of the session.
After the session activities, they were asked about their general thoughts, and they summarized their experience and offered suggestions on how the interface may be modified to improve the user experience.
Finally, after the session, they were asked to complete a post-test survey with both Likert scale questions and open-ended text response questions.
All participants completed this survey less than 1 week after their session.

\begin{figure}[htb]
    \centering
    \includegraphics[width=\linewidth]{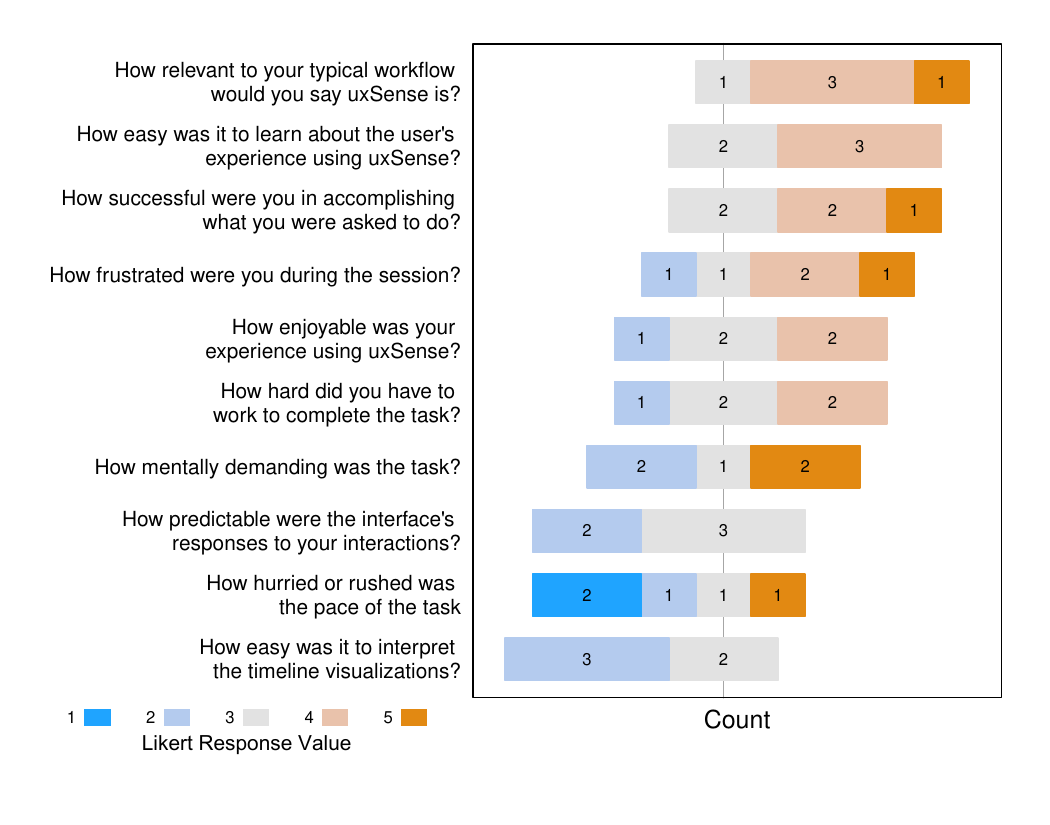}
    \caption{\textbf{Likert responses} to user experience survey
    (1 = strongly disagree, 5 = strongly agree).}
    \label{fig:likerts}
\end{figure}

\section{Expert Review Results}

We originally recruited six UX professionals to participate in our study.
However, the third session failed due to incompatibility problems with the participant's browser, as well as unexpected latency due to the geographical distance between the server and the client coupled with large file size and poor compression. 
For this reason, that participant was unable to complete their session. 
We revised the software based on these experiences and successfully conducted the study with the five participants reported below (numbered 1-6, omitting participant 3). 

\subsection{Think-Aloud Transcripts}

We asked participants to follow a think-aloud protocol during their session.
We used Zoom to automatically transcribe participant utterances, analyzed them, and report our findings below.

In general participants felt that taking notes and marking their corresponding time at the same is demanding. 
However, the multiple timeline interface can relatively reduce some effort. 
For example, participants used the annotation timeline to help them keep track of and organize the notes that they could refer to during their analysis. 
Also, they used the emotion timeline to better understand the user's behavior; for example, P4 commented that \textit{``I found that those positive emotions are related to excitement that he experienced when he [the pictured user] found something really interesting.''}
Although participants found the user's pitch information was indicative of excitement, they felt that the current visualization of pitch information did not help them quickly spot the moments of unusual pitches.
Moreover, participants felt that having access to multiple timelines during their first pass of analysis was overwhelming, because there was too much information to attend to and they wanted to focus on the video. 
Instead, P2 felt the timelines might \textit{``be more helpful in the second round of analysis.''}  
Lastly, participants hoped to be able to configure the layout of different panels, so that they could temporarily hide panels that are non-essential to their analysis at hand. 
They also requested a synchronized video transcript view with the timelines.

\subsection{User Experience Survey}
\label{sec:uxsurveyresponse}

Expert responses to the Likert-scale user experience survey questions described are shown in Figure~\ref{fig:likerts}.
Participants also gave open-ended responses to survey questions; their responses are reported in Table~\ref{tab:survey-qual}.
Participants generally found \systemname{} to be relevant to their typical workflow and allowed them to easily learn about the user's experience, sometimes revealing points of interest in the video that they may have otherwise missed. 
On the other hand, they also found that it sometimes responded in unexpected ways and made them feel frustrated, and that the model output timelines could be difficult to interpret.

\begin{table}[htb]
 \centering 
 \caption{Summary of open-ended survey responses.} 
 \includegraphics[width=\linewidth]{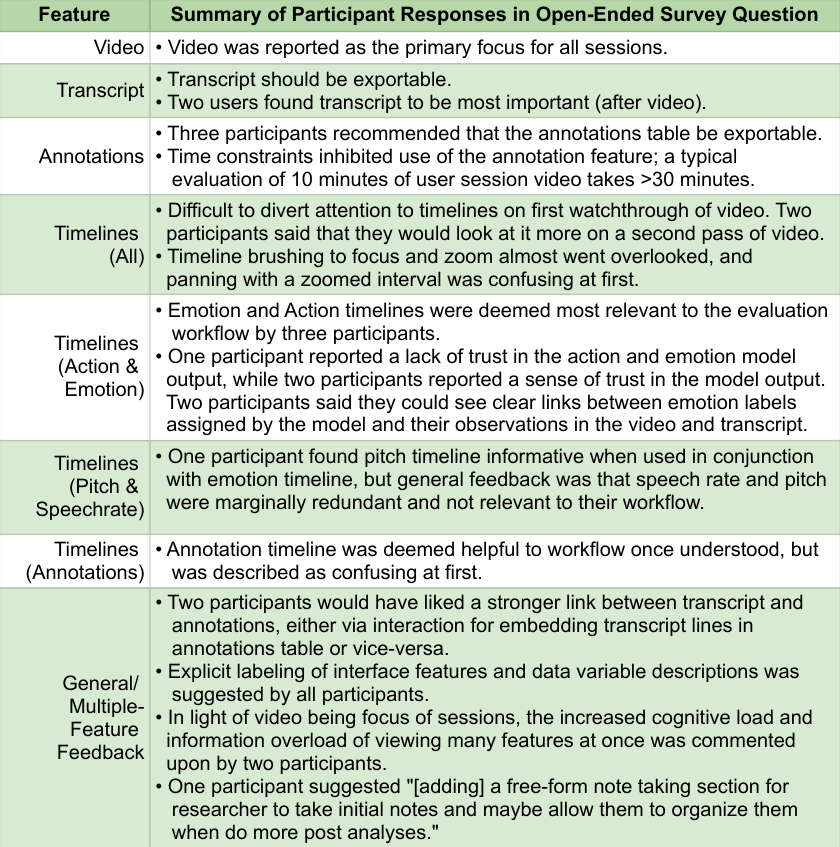}
 \label{tab:survey-qual}
\end{table} 

\begin{figure}[htb]
    \centering
    \includegraphics[width=\linewidth]{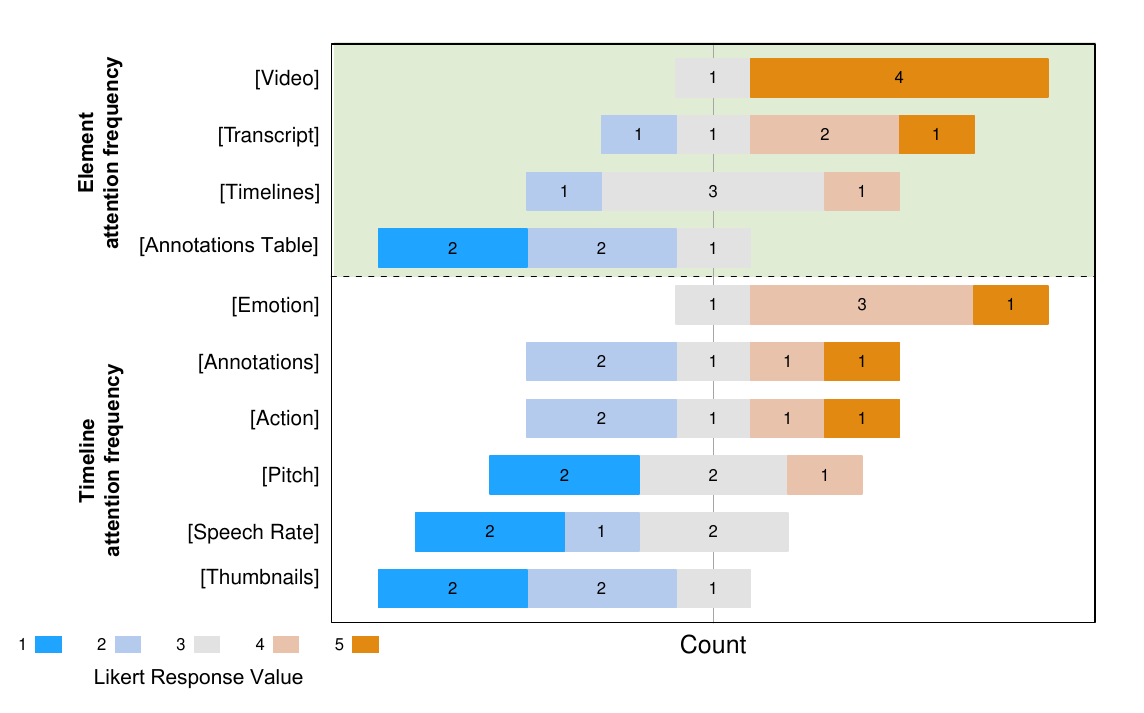}
    \caption{\textbf{Likert responses to time-attention questions.}
    Users were asked to rank time spent on each feature relative to the others (1 = much less time, 5 = much more time).}
    \label{fig:timeuselikerts}
\end{figure}

Their feedback in response to our open-ended survey questions (Table~\ref{tab:survey-qual}) generally suggested design changes that were more in line with universal design guidelines in layout and interaction (e.g., mindfulness of information overload, improved visual feedback to user content creation, and better descriptive labeling of features).
Video playback speed control (0.5x speed, 1.5x speed, 2x speed), 10-second skip-forward/skip-back buttons, and video view resizing were the most commonly requested features, as the video was reported to be the central focus for the participants.   
Most participants also strongly desired a transcript or annotation table export feature, and several participants saw value in visually linking the annotations with the transcripts and/or the selected point or interval on the timelines by embedding two or all three in either the transcript or the annotations table.
It was the combination of these final two points in their feedback that motivated us in our design of the post-study implementation of the ``annotlettes'' feature shown in Sections~\ref{sec:annotlettesystem}~and~\ref{sec:dogfood}---a surprise outcome of our study.

\begin{figure}[htb]
    \centering
    \includegraphics[width=\linewidth]{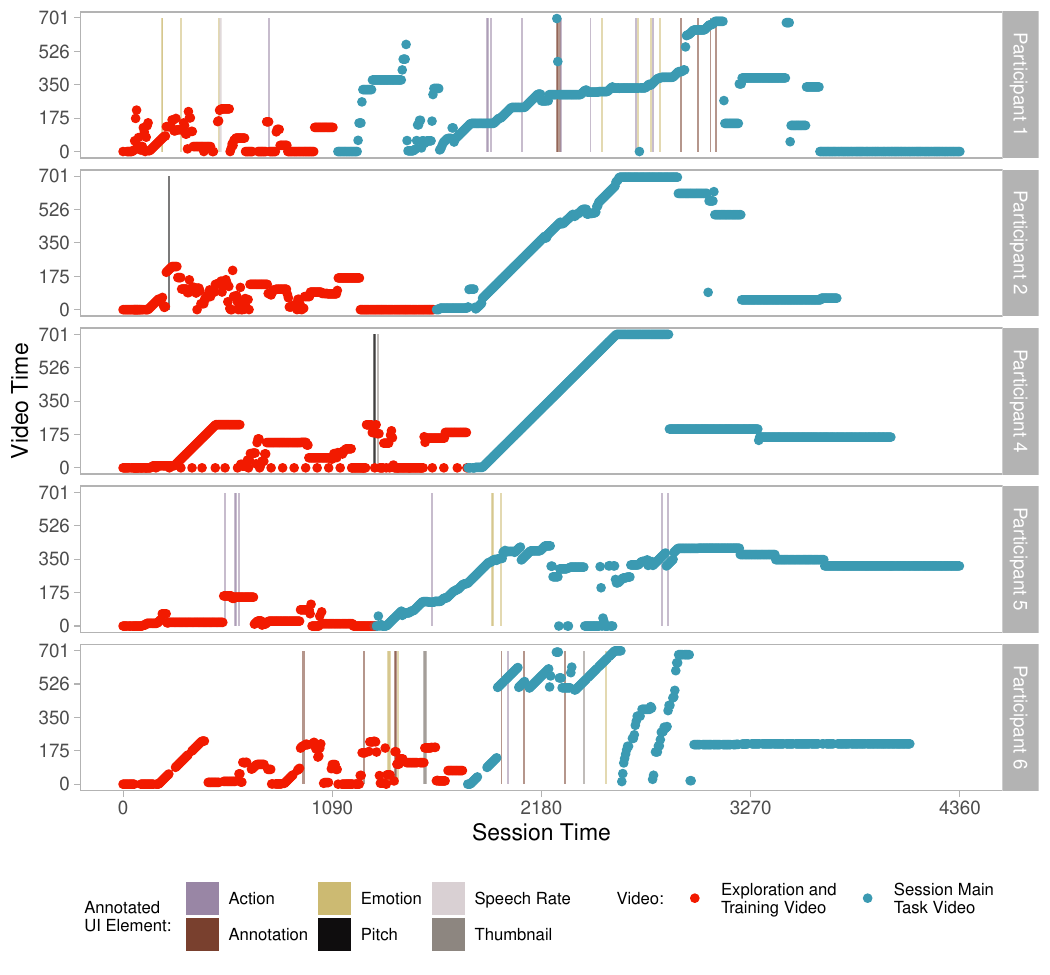}
    \caption{\textbf{Activity logs.} Participant activity in video playbacks.}
    \label{fig:videoplayback}
\end{figure}

\subsection{Time Use: Observed and Self-Reported}

We analyzed event logs, survey responses, and qualitative feedback from user sessions to create a descriptive flow report of their interactions using \systemname{}.
Given the participants' heavy use of the video playback feature, we studied their navigation of the video in detail (Figure~\ref{fig:videoplayback}).
All participants spent a similar share of their time in undirected exploration of features of the interface (shown in red) relative to time spent in the UX evaluation stage of the session.
However, their navigation of the video shows very different viewing patterns during evaluation.
Participants 2 and 4 opted to watch the video the whole way through in a largely linear way before looping back to a small number of points of interest that they spent longer stretches of time on. 
These participants (2 and 4) also scarcely use the annotation feature during the training video, and do not use it at all during the main task.
Participants 1 and 6 start off by skipping around the video in a way that generally progresses forward before looping back and taking a second pass that involves frequent short-range backtracking, and then skip to a small handful of moments of interest. 
These participants (1 and 6) also made the heaviest use of the annotation feature, particularly after their initial skipping around but before their final pass.
Participant 5 performs a lot of short-range backtracking on her first pass, then has a period of frequent skipping back and forth between several intervals in the video that she found most informative before spending longer spans of time examining key points in the video.
Participant 5 makes moderate use of the annotation feature throughout both videos.

Participants were asked to assess where their own attention fell most frequently during their user study as part of the post-session survey (Figure~\ref{fig:timeuselikerts}). 
As we saw in Section~\ref{sec:uxsurveyresponse}, nearly all participants reported spending most of their time examining the video playback, and most reported that the transcript was nearly as central to their evaluation process using \systemname{}.

In order to add depth to our inferences about the participants' use of time during the sessions, we asked them about their own assessment of how and why they distributed their attention during the session.
Participants' self-reported attention time use was recorded on a Likert scale (Figure~\ref{fig:timeuselikerts}).

\subsection{\SYSTEMNAME{} in Three Vignettes}
\label{sec:dogfood}

Based on feedback gathered during the user sessions, the authors extended \systemname{} to produce annotlettes (as described in Section~\ref{sec:annotlettesystem}).
After the sessions in which expert UX researchers used \systemname{} to evaluate a sample user session, we used \systemname{} to evaluate our sessions with them, and then generated annotlettes with our own annotations.
\systemname{} is intended for more detailed reporting than can be contained in a single section, so we demonstrate this feature in three representative vignettes from our evaluation.
While we do not claim that these vignettes are exhaustive, we do think they are representative of participant experiences during our study.

\ornamentbreak

\begin{widequote}
\fontfamily{bch}\selectfont
\begin{figure}[ht]
    \centering
    \includegraphics[width=\linewidth]{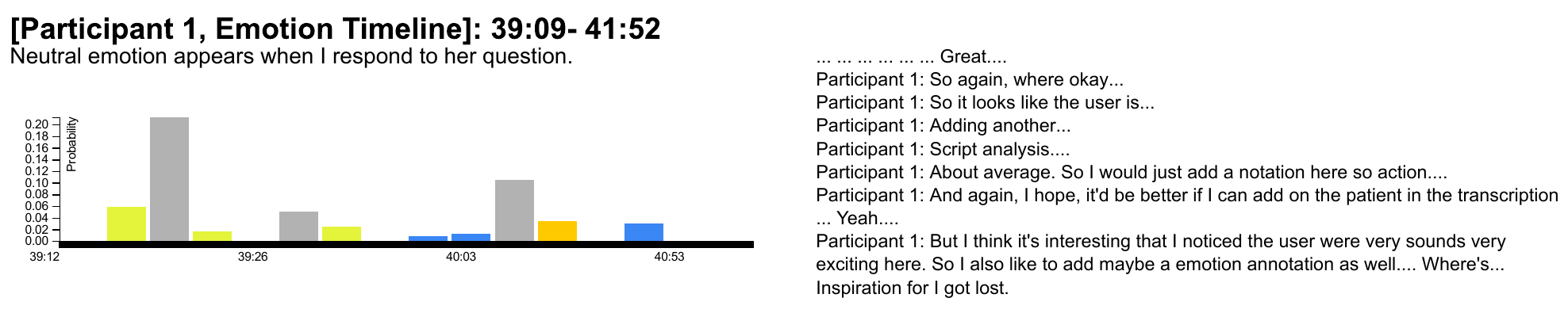}
    \caption{Researcher made an annotation with the observation the appearance of a high-confidence prediction of a neutral facial expression when they responded to the participant's question.}
    \label{fig:annotlet_1a}
\end{figure}

\lettrine[lines=3]{W}{hile}
 exploring the video of Participant 1's user session (video duration: 1 hour and 28 minutes) in \systemname{}, the researcher notices that there were a small number of high-probability ``neutral'' facial expression predictions from the model output that appear as peaks in the emotion timeline (the emotion timeline was directly used by all participants during their user sessions).
 The researcher watches the video from start to until she reaches the first of these peaks, after which point she skips forward directly to the next peak (navigating to a potential point of interest in the video using the emotions timeline was an interaction performed by all participants).
 Using \systemname's focus+context interval selection feature, she adds an interval note (Figure~\ref{fig:annotlet_1a}) highlighting what she observes about the first high-confidence `neutral' label (all participants added at least one interval annotation during their session). 
 Then, navigating to the next high-probability neutral expression in an entirely different phase of the study, she notices that there is another question-answering dialogue between the researcher and the participant about the system, and adds an annotation for that time interval of video as well (Figure~\ref{fig:annotlet_1b}).
 After generating \systemname{} annotlettes, she makes the observation that high-probability neutral expressions tend to follow a low-probability happy expression label (P1 and P6 made interval annotations that explicitly linked patterns in the emotion labels with user behavior).
 Using this pattern observation, she is able to quickly identify other instances of question-answering during the session.
 
\begin{figure}[ht]
    \centering
    \includegraphics[width=\linewidth]{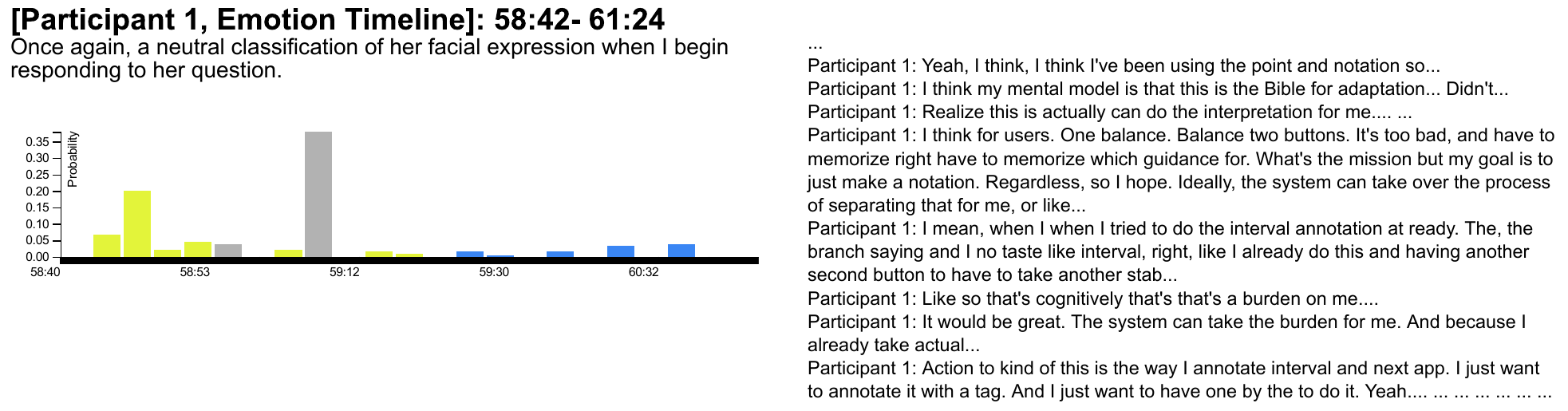}
    \caption{The researcher makes another annotation with the observation of an association between researcher question-answering and the appearance of a high-probability neutral expression.}
    \label{fig:annotlet_1b}
\end{figure}
\end{widequote}

\begin{widequote}
\fontfamily{bch}\selectfont
\begin{figure}[ht]
    \centering
    \includegraphics[width=.55\linewidth]{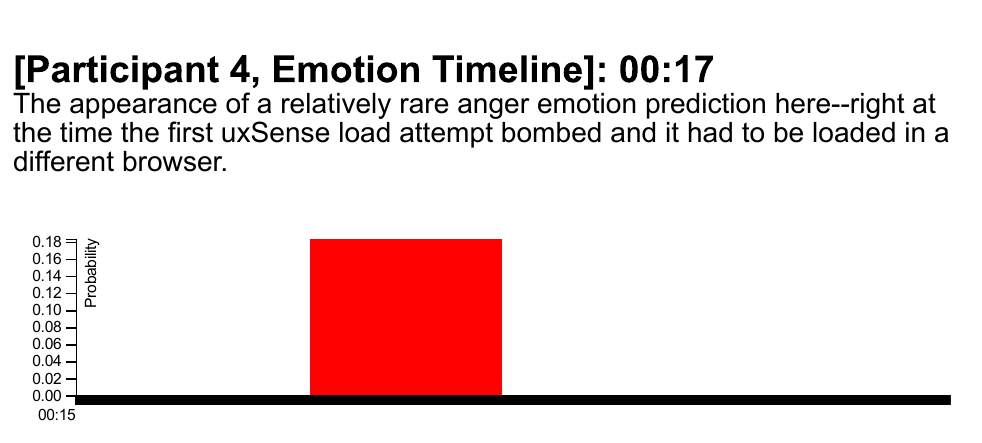}
    \caption{The first appearance of an ``angry'' emotion classification label; here it is associated with the inconvenience of an unanticipated system error that had yet to be worked out.
    There is no associated transcription because there was no dialogue at this specific point in time.}
    \label{fig:annotlet_4a}
\end{figure}

\lettrine[lines=3]{A}{fter}
working her way through user session videos until she reaches  Participant 4's session (video duration: 1 hour and 0 minutes), the researcher notices the appearance of the rare ``angry'' emotion label classification twice (all participants made verbal or annotated observations about the emotions timeline).
Because the anger emotion label has not appeared in her data up to this point, she navigates to these times in the video immediately upon observing them to review the participant's behavior.
She uses the point annotation feature to make a note of what happens during the session associated with the emotion timeline (P1, P5, and P6 made annotations pointing out events in the video and the emotion at the time of the event; in contrast with interval annotations, point annotations tended to appear more frequently and be slightly more detailed than interval annotations). 
The bright red emotion indicator appeared when the user ran into a compatibility issue with the browser she was using, and the system failed to load properly (Figure~\ref{fig:annotlet_4a}).
The researcher seeks out another appearance of the uncommon emotion, again in the same participant's session; this time, it corresponds to an equally uncommon evaluation of a user session: Disapproval toward the user's analytical process (all participants had at least one point during their sessions in which their attention was drawn to labels specifically because they appeared less frequently in the timeline, i.e., to outliers).
She creates another point note describing her observation (Figure~\ref{fig:annotlet_4b}).

\begin{figure}[ht]
    \centering
    \includegraphics[width=\linewidth]{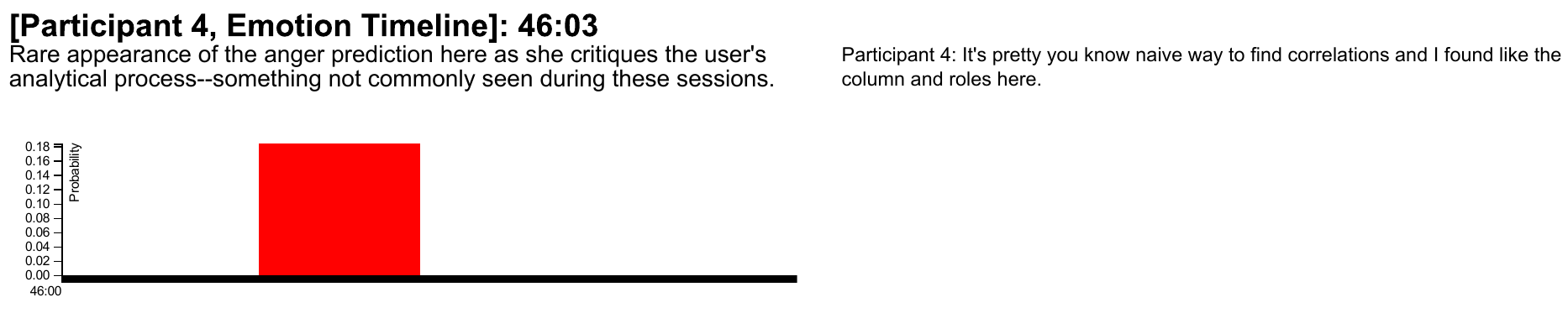}
    \caption{Another rare ``angry'' emotion classification label, again within the same participant's timeline.
    This time it is associated with the equally rare criticism of the user's analytical choice.}
    \label{fig:annotlet_4b}
\end{figure}
\end{widequote}

\begin{widequote}
\fontfamily{bch}\selectfont

\lettrine[lines=3]{M}{oving}
on with her evaluation of the user sessions to Participant 6's session (video duration: 1 hour and 1 minute), the researcher sees that the appearance of action\_290 is something of an outlier in the actions timeline. 
Going directly to that point in the timeline reveals that the participant has said ``I wish I could see this more clearly, because right now it's really small'' in reference to the video playback viewer.
The researcher makes a note of it (P1, P5, and P6 made annotations based on the semantic action label timeline) without looking too closely at the transcript (Figure~\ref{fig:annotlet_6a}).
This particular action is uncommon in this participant's action timeline, so she navigates to its next appearance.
Once again, she sees that the participant is expressing her frustration with the small size of the video playback viewer, and she makes a note of it (Figure~\ref{fig:annotlet_6b}). 
Satisfied for the moment, she generates annotlettes for the annotations she has made thus far, and inserts them as figures in her report. 
Upon inspection of the annotlettes she has created, the researcher is now able to see the semantically-loaded action labels that she tagged during her first pass at evaluating her video dataset.
She finds that action\_290 is ``shaking head;'' she amends her annotation with this new information.
She also sees upon closer review that the actions timeline picked out something that she may have otherwise missed: In the interval selected in the first annotation (Figure~\ref{fig:annotlet_6a}), the participant speaks quietly enough that the transcript failed to capture her complaint about the video playback viewer size.
By using the semantic label visualization, she was able to identify a pattern that would have been missed in an analysis of the transcript alone (P1 and P5 used the actions timeline to support identifying behavior that they did not immediately notice in the transcript).

\begin{figure}[ht]
    \centering
    \includegraphics[width=\linewidth]{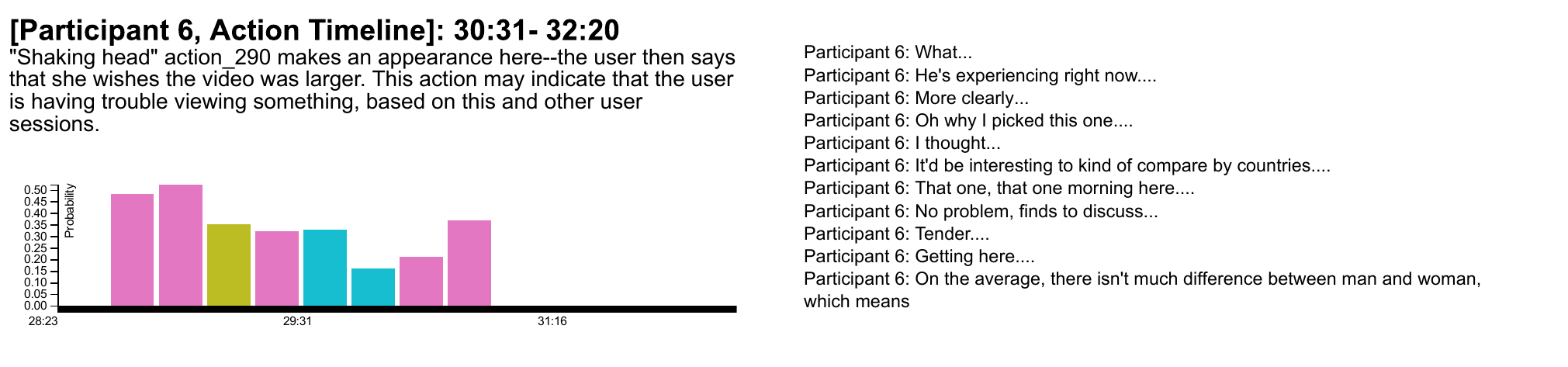}
    \caption{The researcher observes the appearance of action\_290 during a moment in the session in which the participant is expressing frustration at being unable to resize the video playback viewer, but the transcript has not captured her sentiments.}
    \label{fig:annotlet_6a}
\end{figure}

\begin{figure}[ht]
    \centering
    \includegraphics[width=\linewidth]{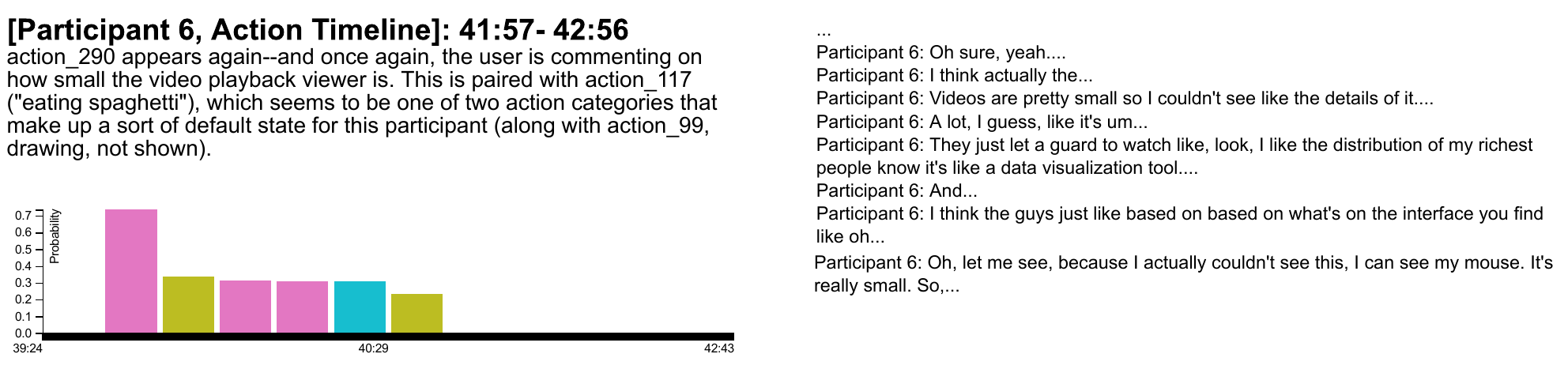}
    \caption{The researcher observes another appearance of action\_290 capturing the participant's complaint about video size, the same design concern featured in Figure~\ref{fig:annotlet_6a}.}
    \label{fig:annotlet_6b}
\end{figure}
\end{widequote}

\ornamentbreak

\section{Proposed System Redesign}
\label{sec:redesign}

Based on our observations and UX expert feedback, we identified some critical changes to the \systemname{} interface.

\begin{itemize}
    \item[A1] \textbf{Annotlettes and data export:} Several users suggested that features that were either generated by the system's backend models or added by the UX researcher, such as the transcript and researcher annotations, should be exportable.
    These requests, coupled with our review of users interactions with their own annotations, resulted in our development of annotlettes (in addition to traditional data export utilities, which we have also added). 
    As noted in Section~\ref{sec:uxsurveyresponse}, the annotlettes feature was one of our post-study additions to \systemname{}. 
    Participants desired a stronger visual link between their annotations, the video, and the data, which annotlettes provide.
    
    \item[A2] \textbf{Video playback features:} All participants wished to be able to speed up, slow down, or resize their view of the video playback.
    For all participants, review of the video was central to their analysis, and there were times during which the screen space taken up by other features of \systemname{} was unnecessary, even if they wound up using a selection of those features at various times throughout their sessions.
    A redesigned system would allow users to hide some or all of the non-video features of the view, in addition to the changes they requested directly.
    
    \item[A3] \textbf{Affordance signifiers:} Some participants found features of the system to not be implied by the interface design.
    In response, we identify several signifiers to \systemname{}'s to make important and easily-overlooked affordances visible: 
    \begin{itemize}
        \item[\adfrightarrowhead] \textbf{Icons} identifying several non-intuitive interactions with the interface, such as the interval selection feature on the zoom focus, and drag directional arrows on the timelines;
        \item[\adfrightarrowhead] \textbf{Tooltips} indicating what the controls in the interface do;
        \item[\adfrightarrowhead] \textbf{Pre-selected interval range on initialization} that highlights the fact that intervals can be selected; and
        \item[\adfrightarrowhead] \textbf{A single annotation button} that is clearly labeled.
    \end{itemize}
    
    \item[A4] \textbf{Session comparison:} The patterns identified in or across user studies are often based on recurring themes or novel outliers that occur amongst, or in contrast across, multiple users.
    The UX system should support qualitative pattern identification across multiple user sessions by allowing the researcher to compare them within and across studies.
    Furthermore, since researchers may be looking for specific patterns, it seemed appropriate to allow them to use their annotations to tag specific segments of video with codes or labels that they create for comparing only those segments across different sessions.
    While this was a design consideration that we discussed prior to our user sessions, it was ultimately not included prior to our study out of respect for the time constraints of our participants and the amount of their valuable time that comparing multiple UX sessions would demand.

\end{itemize}


\section{Discussion}

The use of visualization for analyzing video recording to evaluate human behavior presents opportunities for progress in designing environments that take better advantage of features of the user, their surroundings, or their session, and respond accordingly.
However, there are ethical considerations and limitations in using video data.

While our system may have faced a few design and performance issues during the expert user study, the results were largely affirming of the relevance of, interest in, and frankly, need for ML-supported HCI.
Equally importantly, our participants directly and indirectly highlighted improvements that we have implemented in our redesign of the system described in Section~\ref{sec:redesign}.

Perhaps our favorite result from the study was in finding a way to not only use the system to evaluate itself, but in using that self-evaluation to extend \systemname{} so that it could aid in writing narrative vignettes that highlighted important moments for further design revisions. 
This represents, to us, the unlimited potential for growth and improvement in ML-HCI-VIS evaluation systems.

\subsection{Beyond Desktop UX}
\label{sec:beyond-desktop}

While we so far have explored the use of \systemname{} for desktop computer applications, it is clear that the same ideas can be applied to other forms of computing, such as immersive 3D and mobile computing.
In fact, one of our original motivations for \systemname{} was to support 3D immersive analytics evaluation.
This would, for example, enable extending tools such as ReLive~\cite{DBLP:conf/chi/HubenschmidWFBZ22} to include automatically derived metrics.
It would require adding the following features to the data model:

\begin{enumerate}

    \item[\faIcon{map-marked-alt}]\textbf{F4 - Physical navigation:} 
    User physical location over time.

    \item[\faIcon{hands}]\textbf{F5 - Limb tracking:}
    The ability to track both fine (fingers and hands) and gross (arms, legs, torso, head) motor interaction.
    
\end{enumerate}

Several of our filters (Table~\ref{tab:filters}) already support such data, including for detecting 3D pose~\cite{Pavllo2019}, 3D joint angles~\cite{Batch2018gesture}, and actions~\cite{Matteson2014}.
However, we leave such extensions to future work.

\subsection{Limitations}

There are several limitations with the technical approach we propose in this paper.
Even state-of-the-art computer vision and machine learning algorithms are still far from perfect in accurately detecting the features that \systemname{} visualizes, and thus there is significant noise in the process.
The fact that our participants mostly relied on the raw video footage rather than the extracted feature timelines during our expert review may be an indication that the noise was too overwhelming to rely upon.
In a sense, this could be seen as a negative result: perhaps automated or even semi-automated UX analysis is doomed to failure?

We offer several answers to this question.
On the human side of the equation, it is possible that this bias in favor of raw video is merely a habitual effect arising from our participants' everyday practices as professional UX designers.
It could also be due to the limited amount of time that UX evaluators were exposed to the tool, as suggested by Fan et al.\ who observed similar behaviors of UX evaluators in the evaluation of their UX video analytics tool \cite{fan2019vista}.
Thus, we suspect that this effect will probably decrease as UX evaluators embrace this kind of automated tool in their daily workflow over time. 
On the technology side, the noise in our feature streams will likely reduce over time as computer vision and machine learning technologies improve.
Furthermore, the problem can be somewhat mitigated by offering several complementary data streams that serve to provide an accurate overview despite inaccuracies in individual streams.
However, there is nothing to prevent a user from importing manually annotated data into \systemname{} as a complement.
Furthermore, \systemname{} also gives the user access to the original video and audio data, thus allowing them to verify (and correct) any automatic annotation in the interface.

The other side of this coin is that automatic metrics may lead to overreliance, anchoring effects, and even a decay in analytical reasoning.
Furthermore, the lack of transparency for many of the computer vision models may certainly impact the user's trust in the system.
Involving the human in the loop and triangulating multiple metrics may mitigate these dangers; however, it remains a cautionary tale when employing automatic methods. Alternatively, recent research showed that when and how automatic metrics are revealed to UX evaluators can affect their trust and performance \cite{Fan2022Human-AI}. Thus, another possible direction is to investigate human-AI collaborative approaches.

In terms of limitations of our research methodology, our evaluation was done with a very small set of professional UX researchers from tech companies. 
We only involved a single sample user study session video for the evaluation activity. 
Furthermore, the qualitative nature of our study provides few quantitative measurements on performance and the lack of a baseline makes comparison to other tools impossible.
We made these choices with the intention of collecting actionable feedback for improving \systemname{} further rather than summatively assessing its utility.
However, we acknowledge that our single evaluation is not generally representative of the many potential variations of products and end-users, and that further studies are needed in the future.

Part of our evaluation methodology is based on using \systemname{} itself to evaluate \systemname{}.
There is obviously a risk in this practice because missing features in \systemname{} would lead to these aspects of the system not being evaluated; an intrinsic ``blind spot'' in our methodology.
We have mitigated this risk by triangulating the findings with our own experience as visualization and UX researchers, akin to how we envision UX designers will use \systemname{} in practice.

Finally, in choosing the commercial data analysis and visualization tool Tableau as the platform of choice for our expert review, we restricted the usability study session used as a dataset to a mature software package with a polished user interface.
This limited the scope and scale of usability issues that our UX designer participants could find in the user study data.
In the future, it would be interesting to perform a follow-up study involving a more experimental and early-stage user interface, or for a more academic user study conducted in a controlled laboratory setting.

\subsection{Ethical Considerations of Video}

In today's surveillance society, an increasing portion of our lives takes place inside a camera viewfinder.
It seems as if every week uncovers some new horror of how digital video can be abused to threaten the privacy, security, and even safety of people just trying to live their lives.
Thus, it could well be argued that a system such as \systemname{} is misguided in that it builds on fundamentally problematic ideas about recording human participants, and could even facilitate future abuses in the same vein. 
Indeed, while our current prototype does not include a facial recognition component, we could easily see the utility of having such a filter for when a system is deployed in the field and the researchers want to track how specific people use the system over time.

Rather than merely disavow any future such events as beyond our control, let us here acknowledge that this is a possible outcome.
We ourselves commit to safeguarding our own approach and prototype so that the recordings are not distributed or used for other purposes than for what participants gave informed consent to.

We also note that researchers already collect plenty of video recordings of their participants, much of which is only lightly analyzed and then archived.
These videos will obviously identify the participants.
Due to the prohibitive size of video, we suspect that much of this data resides on unprotected network or external drives.
However, a fundamental feature of our approach is to process high-bandwidth video data to extract key data streams from the footage.
These extracted data streams are refined and precise; the position of a person in three dimensions, their fatigue level, or the direction of their gaze.
After all relevant data has been extracted, the original video can be deleted, thereby saving storage space and eliminating identifiable likenesses of participants.
Thus, it could be argued that our approach may actually improve privacy, as it allows researchers to safely discard video while retaining deidentified data.

\section{Conclusion and Future Work}

We have proposed a visual analytics tool called \systemname{} for visualizing multiple timelines of data streams extracted from video and audio recordings of usability sessions.
We then present results from an expert review where UX professionals used the tool to understand a usability session conducted using Tableau.
Finally, we used \systemname{} itself to analyze these expert review sessions.

Our prototype only includes a small set of filters to demonstrate our concept, and we can easily see the need for many additional filters.
Fortunately, all feature extraction filters in \systemname{} are Open Source modules, and we anticipate generously extending this library of filters to include other Open Source modules.
Furthermore, it is common for UX professionals to analyze multiple sessions from different users to determine common issues.
\systemname{} is currently restricted to evaluating a single user's session in a post-hoc manner.
In the future, UX professionals may need to analyze user experience across multiple users, perhaps even real-time.

In fact, many interactive systems now consist of multiple, connected, and collaborating devices, such as in smart rooms, for groups of smartphones working together, or for an individual user's menagerie of personal devices.
By taking advantage of this networked system of many devices, we can extend \systemname{} and develop new systems like it, and step a little closer to HCI's dream of context-aware and ubiquitous computing~\cite{Weiser1991}.

\section*{Acknowledgments}

This work was partially supported by the U.S.\ National Science Foundation grant IIS-1908605 and the Natural Sciences and Engineering Research Council of Canada through the Discovery Grant program.
Any opinions, findings, and conclusions or recommendations expressed here are those of the authors and do not necessarily reflect the views of the funding agencies.

\bibliographystyle{plainnat}
\bibliography{uxsense}

\end{document}